\documentclass[preprint,aps]{revtex4}
\usepackage{graphicx}
\usepackage{bm}
\usepackage{color}
\usepackage{amssymb}
\usepackage{epsfig}

\begin{document}

\title{Random matrix ensembles with column/row constraints. II}
\author{Suchetana Sadhukhan and  Pragya Shukla}
\affiliation{Department of Physics, Indian Institute of Technology,
Kharagpur, India}

\date{\today}


\begin{abstract}

  We numerically analyze the random matrix ensembles of real-symmetric matrices with column/row constraints for many system conditions e.g. disorder type, matrix-size and basis-connectivity. The   results reveal a rich behavior hidden beneath the spectral statistics 
and also confirm our analytical predictions, presented in part I of this paper,  about the analogy of their spectral fluctuations with those of a critical Brownian ensemble which appears between Poisson and Gaussian orthogonal ensemble.

\end{abstract}


\maketitle

\section{Introduction}

Random matrices with column/row constraints appear as the mathematical representations of complex systems in diverse areas e.g bosonic Hamiltonians such as  phonons, and spin-waves in Heisenberg and XY ferromagnets, antiferromagnets, and spin-glasses,  euclidean random matrices, random reactance networks, financial systems and Internet related Google matrix  etc (see section II and references in \cite{ss}).  It is therefore relevant to seek the information 
about their  statistical behavior; this not only helps to probe the related complex system but also in revealing the hidden connections among seemingly different complex systems. 

 The  standard route for the statistical analysis of a random matrix ensemble is based on its eigenvalues/ eigenfunctions fluctuations. This in turn requires  a knowledge of the joint probability distribution function (JPDF) of the eigenvalues and eigenfunctions. An integration of JPDF over undesired variables then, in principle, leads to the fluctuation measures for the  remaining variables; for example, the spectral fluctuation measures can be obtained by an integration over all eigenfunctions.  As discussed in \cite{ss}, the JPDF for the column/row constrained ensembles can be derived from the ensemble density but its further integration over eigenfunctions is not an easy task. This motivates us to seek, an alternative route i.e the numerical investigation of the ensembles; our  focus in this paper is on following objectives: 

(i) to understand the effect of randomness (both type and strength) and basis-connectivity on the fluctuation measures.

(ii)  to analyze the size-dependence of the  fluctuation measures, 

(iii)  search for  the existence of any universal features in the statistical behavior.

The motivation for the  first objective comes from previous studies of the  systems 
with Goldstone symmetry; the column constraints in the matrix representation of their operators  originate from the invariance under a uniform shift \cite{ss}.  In absence of disorder, the low frequency excitation in these 
systems have long wavelengths and are described by the equations of motion based on 
macroscopic properties. Disorder causes local fluctuations in the properties but 
excitations are believed to couple only to their averages over a volume with linear dimensions set by a wavelength. Consequently, higher dimensions, where averaging is more effective, are expected to be disorder-insensitive \cite{gc}. But almost all these results are based on average behavior of the spectral density and inverse participation ratio (IPR). As the statistical properties can in principle be expressed in terms of the eigenvalues/eigenfunctions fluctuations of the related operators, it is natural to query whether disorder has any, and if so, what impact on the fluctuations of the excitations, especially in higher dimensions.  As mentioned in section 2 of \cite{ss}, the spectral statistics of the low lying bosonic excitations above a ground state $E_0$  is analogous to that of an ensemble of column constrained matrices with column constant $\alpha_l=E_0$ (for $l=1,2,..N$). The latter being statistically similar  to the case $\alpha_l=0$ (except for a shift in the spectrum), the spectral information about Goldstone modes can be obtained by an analysis of the $\alpha_l=0$ ensemble.  Keeping this in view, the present analysis is confined  to the zero-column constraint ensembles.

The modelling of  complex systems by random matrix ensembles usually requires consideration of the infinite size limit. The numerical analysis however involves only finite size matrices. To translate the numerical information to real systems, it is  necessary to analyze finite size effects which forms the basis of our 2nd objective.

The necessity for the third objective originates from our recent study \cite{ss} of the JPDF of the column constrained ensembles (CCE) for two different disorder types and basis-connectvities. Our results  indicated a mathematical analogy of their JPDF with that of a Brownian ensemble (BE) appearing between Poisson and Gaussian orthogonal ensemble (GOE); this in turn implies a CCE-BE analogy  for the local spectral fluctuation measures. The implication is seemingly in contrast to a previous study \cite{yvf} which predicts the analogy of the two point level-density correlation of CCE to  that of a GOE.  But, as explained in \cite{ss}, the  BE-statistics in an energy range near $e=0$ is indeed of GOE type (see the discussion in 2nd paragraph after eq.(73) in \cite{ss} ) and, therefore, our prediction agrees with \cite{yvf} near $e=0$.
For other energy ranges,  a direct  numerical comparison of their fluctuation measures is needed to reconfirm the  CCE-BE analogy.  (Although a comparison of CCE numerics with analytical BE-results  would have been more desirable but, due to  technical complexity related to the orthogonal-space integration, available BE-results are mostly asymptotic).

To fulfill the above objectives, here we numerically study  CCEs for many system conditions. More specifically, we consider the Gaussian and bimodal distributions of the off-diagonal matrix elements for many disorder strengths and system sizes; the diagonals are obtained by invoking the column sum rule. The dimensionality in the matrices is taken into account by varying basis-connectivity i.e by choosing the matrix elements forming nearest-neighbor pairs non-zero and all others zero (still subjected to column sum rule).

The paper is organized as follows. 
Section 2 presents the ensemble densities  for the cases considered in our numerical analysis; these ensembles are special cases of those considered in section 4 of \cite{ss}.  
An analysis of the spectral fluctuation measures requires a specific rescaling of the spectrum which in turn depends on whether the spectrum is ergodic or not; this is discussed in section 2.A alongwith a numerical study of the ergodicity for CCEs, BEs as well as GOE; the latter is included to illustrate the deviation of CCEs from GOE. Sections 2.B and 2.C present the  numerical results for the average behavior and fluctuations, respectively,  of  CCEs with Gaussian and bimodal distributed off-diagonals, both cases considered for infinite as well as finite basis-connectivity and many matrix sizes; the objective of these sections is not only to verify the theoretical insight derived in \cite{ss} but also to  provide information about  mathematically intractable fluctuation measures.  The results of a numerical comparison of the CCE-BE fluctuations are given in section 3 which reconfirm our theoretical predictions of \cite{ss}. Section 4 concludes this study with a brief reviews of our main results and open questions.  


\section{column constrained ensembles: four types}.

To investigate the sensitivity of the statistical fluctuations to the nature of randomness  and basis-connectivity, we consider following CCE  ensembles of real-symmetric matrices $H$ with  $\rho(H)$ as their  probability densities:

\vspace{0.1in}

\noindent{\it (a)  independent off-diagonals with Gaussian distribution (infinite range or $d=\infty$ CCGE case)}
\begin{eqnarray}
\rho(H) &=& \mathcal{N} \;   {\rm exp}\left[-\gamma \sum_{k,l; k\not= l} H^2_{kl} \right] \; \prod_{l=1}^N \delta \left(\sum_k  H_{kl} \right)
\label{rho2}
\end{eqnarray}
This case corresponds to case II of \cite{ss} with $ \alpha=0$.

\vspace{0.1in}

\noindent{\it (b)  Independent,  nearest-neighbor Gaussian hopping in $d$-dimension (finite $d$ CCGE case)}

\begin{eqnarray}
\rho(H) 
&=&\mathcal{N} \;   {\rm exp}\left[-\gamma \sum_{k=1}^N \sum_{l; l \in z(k)} \; H_{kl}^2 \right] \; \prod_{k, l \not\in z(k)} \delta(H_{kl}) \; \prod_{l=1}^N \delta \left(\sum_k  H_{kl} \right)
\label{rho3}
\end{eqnarray}
with $z$ as the number of nearest neighbors of a site. The symbol $\sum_{l \in z(k)}$ refers to a sum over all nearest-neighbors of a given site $k$ (excluding $l=k$ site). This case corresponds to case III of \cite{ss} with $ b_0=b_1=0, \alpha=0$.

\vspace{0.1in}

\noindent{\it (c)  Independent off-diagonals with bimodal distribution } (infinite range or  $d=\infty$ CCBE case). 
\begin{eqnarray}
\rho(H) =\left( \prod_{k,l;  k < l}^{N} \left[ \delta(H_{kl} -1) + \delta(H_{kl} + 1) \right] \right) \; \prod_{l=1}^N \delta \left(\sum_k  H_{kl} \right)
\label{birho1}.
\end{eqnarray}
This case corresponds to case IV of \cite{ss} with $a=1, \alpha=0$.

\vspace{0.1in}

\noindent{\it  (d) Independent, nearest-neighbor bimodal hopping in $d$-dimension } (finite-$d$ CCBE case)
\begin{eqnarray}
\rho(H) = \mathcal{N} \; \left(\prod_{k,l;  l \in z(k) }^{N} \left[ \sum_{q=\pm 1} \delta(H_{kl} - q a ) \right] \right) \; 
\left(\prod_{k,l; l \not\in z(k)}^N \delta \left( H_{kl}  \right) \right) \; 
 \prod_{l=1}^N \delta \left(\sum_k  H_{kl} \right)
\nonumber \\
\label{birho2}.
\end{eqnarray}
This case corresponds to case V of \cite{ss} with $a=1, b_1=0, \alpha=0$.

For numerical computation 
of the eigenvalues and eigenfunctions of each of the ensembles above, we exactly diagonalize thousands of matrices of  sizes $N=512, 1000, 2197, 4913$ using LAPACK subroutines; by choosing an appropriate number of matrices, the statistical data size for each $N$ is ensured to be same. 
Further, as clear from eqs.(\ref{rho2},\ref{rho3},\ref{birho1},\ref{birho2}), the spectral statistics for these cases is independent of the off-diagonal variance (a rescaling of the matrix elements by the variance leaves the statistics unaffected); this is also confirmed by our numerics (shown here for the level-density case in figure 1). For fluctuations studies, the variance of non-zero off-diagonals in eqs.(\ref{rho2},\ref{rho3}) is therefore kept fixed at $\gamma=1/2$.

Prior to calculation of the fluctuation measures, it is necessary to determine the ergodic as well as stationary aspects of the spectrum.

\subsection{Ergodicity and stationarity of the spectrum }

The spectral density $\rho_e(e)$ at a spectral scale $e$ is defined as 
$\rho_e(e)= \sum_{k=1}^N \delta(e-e_k)$. Under assumptions that $\rho_e$  consists of two spectral scales, a slowly varying component $\rho_{sm}$, often referred as the secular behavior, and a rapidly varying component, $\rho_{fluc}$, one can write $\rho_e =\rho_{sm} + \rho_{fluc}$ (also referred as the local  random excitations of $\rho_{sm}$) \cite{brody, haak}.  A spectral averaging over a scale larger than that of 
the  fluctuations gives $ \int_{e-\Delta e/2}^{e+\Delta e/2} {\rm d} e \; \rho_{fluc}(e) =0$ and  $\rho_{sm} = \frac{1}{\Delta e} \int_{e-\Delta e/2}^{e+\Delta e/2} {\rm d} e \; \rho_e(e)$.  Previous studies of a wide range of complex systems indicate $\rho_{sm}$ to be more system specific as compared to the fluctuations; the latter reveal a great deal of universal features and often depend only on the global constraints of the system.  The information about  local fluctuations can therefore be better revealed by an "unfolding" the spectrum i.e  removing its system-dependent part $\rho_{sm}$ by a rescaling  $r_n=\int_{-\infty}^{e_n} {\rho_{sm}}(x) \; {\rm d}x$ of the eigenvalues $e_n$ which results in uniform local mean spectral density \cite{haak,brody}.



Experimental studies often analyze the behavior of  a single spectrum which leaves the unfolding by $\rho_{sm}$  the only option. But theoretical modelling of a complex system by a random matrix ensemble is based on the assumptions of ergodicity of global and local properties. As discussed in section 2.2.3 of \cite{bg}, the ergodicity of a global property such as the level density  
implies $\rho_{sm}(e)=R_1(e)$, with $R_1(e)$ as the ensemble averaged level density, and one can use $R_1(e)$ as a substitute for $\rho_{sm}(e)$ for various analytical purposes. It is therefore necessary to check the ergodicity of the level density which 
requires \cite{haak, brody, bg} 
\begin{eqnarray}
\langle \rho_{sm}(e) ^2 \rangle - \langle \rho_{sm}(e) \rangle^2 \rightarrow 0,  \qquad \qquad \langle \rho_{sm}(e) \rangle= R_1(e)
\label{erg}
\end{eqnarray}
with $\langle . \rangle$ implying an ensemble average for a fixed $e$.
A comparison of the ensemble and the spectral averaging for the  spectral density of CCGE, BE as well as GOE is shown in figure 1; here each case is considered for an ensemble of $5 \times 10^3$ matrices of size $N=1000$. As seen from the figure 1(d),  $\rho_{sm}(e)$ even for a single GOE matrix almost coincides with $\langle \rho_{sm}(e)\rangle$ as well as $R_1(e)$ for all $e$, reconfirming its ergodicity. For CGGE and BE, the $e$-dependence of the $\rho_{sm}$ fluctuates (weakly) from one matrix to the other  but the ensemble averaged  $\rho_{sm}(e)$ does agree well with $R_1(e)$.
Although  the figure 1  displays the result only for one system size, we find similar analogy for higher $N$ values too, with deviation from one matrix realization to other diminishing with higher $N$. This indicates an approach to ergodicity for the level density in large $N$ limit.


%

%
%

An ergodic level density however does not by itself implies the ergodicity of the local density fluctuations. 
As discussed in section 2.2.3 of \cite{bg}, the ergodicity   of a local property say "$f$" can be defined as follows: let $P_e(s)$ be the distribution function of $f$ in the region "$e$" over the ensemble and ${\tilde P}_k(s)$ be the distribution function of $f$ over a single realization, say "$k$" of the ensemble. $f$ is called ergodic if it satisfies the following conditions:
(i) $P_e(s)$ is stationary i.e $P_e(s) = P_{e'}(s)$  for arbitrary $e, e'$,
(ii) for almost all $k$, ${\tilde P}_k(s)$ is independent of $k$ i.e ${\tilde P}_k(s) = {\tilde P}(s)$, 
(iii) $P_e(s) = {\tilde P}(s)$.
However if condition (i) is not fulfilled,  $f$ is then referred as locally ergodic. More clearly, an ensemble is locally ergodic in a spectral range say $\Delta e$ around $e$, if the averages of its local properties over the range for a single matrix  (referred as the spectral average) are same as the averages over the ensemble  at a  fixed $e$ (referred as the ensemble average). To analyze the ergodicity of local spectral fluctuations, we consider  a standard measure, often used for this purpose, namely,  {\it the width $\sigma(k,e)$ of the distribution $P(k;s,e)$ of the eigenvalue spacings $S_k$ between $k^{th}$ neighbors ($k=0,1,2..$) within a range $(e-\Delta e/2, e+\Delta e/2)$ with $s=S_k/D$ and $D$ as the mean spacing} \cite{bg, brody}; (note the mean of the distribution is $k+1$). 
Figure 2 compares the ensemble averaged $\sigma_k$ with the spectral average for  CCGE, BE and GOE, separately for each case (using eq.(35) and eq.(39) of \cite{bg} for the spectral and the ensemble average, respectively)). As the figure indicates, contrary to GOE, the local fluctuations in both CCGE as well as BE are non-ergodic and their unfolding with $R_1(e)$ for finite $N$ cases is not appropriate (as it may introduce spurious fluctuations).  It is therefore necessary to determine the local spectral density $\rho_{sm}$ either by theoretical, experimental or computational route.

As mentioned above, the stationarity is an important aspect  which implies an invariance of the local fluctuation measures along the spectrum-axis. A non-stationary spectrum can at best be locally ergodic and therefore a complete information about the spectrum requires its analysis at various spectral ranges.  Figure 3 displays the behavior of an ensemble averaged $\sigma(k;e)$ for CCGE, BE and GOE  for many $k$-values and for a few spectral ranges taken from the region between edge and bulk.  Note, as already known \cite{me}, GOE is also not exactly stationary  and the edge statistics shows significant deviation from that of the bulk (our numerics shown in figure 3(d) reconfirms this behavior). As clear from the figure, the regions in which translational invariance holds for CCGE and  BE are much smaller than in GOE.

\subsection{Spectral density and inverse participation ratio: average behavior}

 The average spectral density  $\rho_{sm}$ is 
a standard measure  for experimental or computational  studies of many complex systems (often based on single system analysis); 
figures 4,5  depict $\rho_{sm}$ behavior of a column constrained matrix, for two disorder types, namely, Gaussian and bimodal, for various dimensions (i.e basis connectivity)
and  disorder strengths.  The effect of dimensionality on the average 
behavior is clearly visible from these figures. As shown for the Gaussian case in figure 4, the dependence of $\rho_{sm}$ on the disorder strength $\gamma$ can be scaled out.  The size $N$ has no effect on $\rho_{sm}$ for short-range basis connectivity 
(nearest-neighbor hopping cases with low-dimensionality  $d \leq 3$), for both type of disorders (ensemble(\ref{rho3},\ref{birho2})) . The numerics for different $N$  for the  long-range basis-connectivity (ensemble (\ref{rho2},\ref{birho1})), displayed in figure 5(c),  however indicates  following $N$-dependence  $\rho_{sm}(e)=\frac{1}{\sqrt{N}} f\left(\frac{e}{\sqrt{N}}\right)$. A comparison shown in figure 12(a)  suggests  the bulk behavior of the function $f$ as semicircle :  $\rho_{sm}(e)=\frac{1}{\pi a}\sqrt{2a N -e^2}$ with $a$ a constant. An important point to note from figure 5(c) is the same $N$ dependence of $\rho_{sm}$ in both edge and bulk; the critical BE with $\mu \propto N$ is  expected to model the  $d=\infty$ CCE-statistics in both energy regimes (see section V.B of \cite{ss} for the details).      
Furthermore the edge-behavior seems to be in agreement with the level-density of Goldstone modes in spin-glasses for $d \ge 3$: $\rho_{sm}(\omega) \sim \omega^{q}$ where $\omega=e-e_0$ and $q=2$ for $d=3$ and $q=1.5$ for $d=\infty$. (Based on various studies, the behavior of level density for Goldstone modes  in spin glasses is believed to be as follows: $\rho_{sm}(\omega) \propto \omega^{-1/3}$ for $d=1$,  $\rho_{sm}(\omega) \propto \omega^{2}$ for $d=3$ \cite{gc,stinch} and  $\rho_{sm}(\omega) \propto \omega^{3/2}$ for  infinite-range spin glass \cite{gc,bray}).
Here, for clarity of presentation, the behavior for $R_1(e)$ and $\langle \rho_{sm}(e) \rangle$ are not included in figures 4,5 but, based on figure 1, their dependence on disorder and system-size is expected to be same as that of $\rho_{sm}$.  
 
As discussed in \cite{ss}, a typical eigenvector of a real-symmetric column constrained matrix displays a new characteristic: it is the lack of single basis-state localization  (the minimum number of basis-states required for  localization is two) which is in constrast with an unconstrained real-symmetric matrix. This tendency can be better analyzed by a numerical study  of the inverse participation ratio (IPR) $I_2$, the standard tool to describe the localization behavior of an eigenfunction.   It is defined as $I_2(O_n)= \sum_{k=1}^N  |O_{kn}|^4$ for an eigenfunction $O_n$ with $O_{kn}$ as its components in a $N$-dimensional basis. The minimum number of basis-states available for localization restricts the maximum IPR $ I_2 \le \frac{\kappa}{2}$ with $\kappa=1$ or $2$ for a typical  eigenvector of a matrix with or without column constraints, respectively \cite{ss}. As can be seen from figure 6, the maximum IPR indeed does not go above 0.5, except in figure 6(f) where it reaches 0.6; the latter seems to be an edge effect.

The localization length in general fluctuates from one eigenfunction to the other.  The average localization behavior of the eigenfunctions within a given spectral range can be given by a spectral averaged $I_{2}$ (i.e average of IPR of all eigenfunctions within the range) \cite{ravin}.   
In case of an ensemble, it is useful to consider the  ensemble as well as spectral averaged $I_{2}$, later referred simply as $\langle I_{2}\rangle$. Figure 6 indicates the existence of a scale-invariance of  $\langle I_{2}\rangle$ with respect to off-diagonal variance and therefore degree of disorder for both types. As clear from the insets in figure 6, 
the  localization tendency for the low lying eigenfunctions (lower energy excitations) for the Gaussian and Bimodal types is just the reverse  for short-range basis 
connectivity but similar for the long range. The fits in the insets for each case 
suggest $ \langle I_{2,sm}\rangle \sim \omega^{1/3}$ for $d=1$, $\langle I_{2,sm}\rangle \sim \omega^2$ for $d \ge 3$ for small-$\omega$. 
Previous spin glass studies on the Goldstone modes give the avearge (spectral) localization length 
$\zeta(\omega) \propto \omega^{-1/3}$ for $d=1$ and $ \zeta(\omega) \propto \omega^{-2/3}$ 
for $d \ge 3$ \cite{gc} which implies, using $I_{2,sm} \propto \zeta^{-d}$ \cite{ravin}, 
$I_2 \propto \omega^{1/3}$ for $d=1$ and $I_{2,sm} \propto \omega^2$ for $d \ge 3$.   
As the fits in the inset indicate, the CCE-results are in agreement with spin glass behavior.

\subsection{Spectral fluctuations}

Our next step is to probe the local fluctuations around the spectral density for  two disorder-types and for different connectivity in the lattice. 
As discussed above, $\rho_{sm}(e)$ for CCE is  $e$-dependent and a comparison of local fluctuations from two different spectral regions with different $\rho_{sm}(e)$ requires an unfolding of the spectrum. 
In past, many ways for unfolding have been suggested e.g.  polynomial unfolding (approximating the empirical staircase function by a polynomial) \cite{am}, Gaussian unfolding (Gaussian smoothing of each delta peak in $\rho_e$) \cite{haak,gmrr}, Fourier unfolding (removal of short scale wavelengths using Fourier transformation) or local unfolding (local mean  level density assumed to be constant within  a narrow spectral-window) \cite{gmrr}, detrending the fluctuations \cite{mlsf} etc. 
Previous studies of many systems indicate the statistics in the low-density regions e.g. edge to be sensitive to the  procedure used but that in the bulk is unaffected. Although  Gaussian unfolding is often believed to be suitable in the edge region, the results are sensitive to  width of the Gaussian smoothing \cite{gmrr}. Our numerical experiment with various unfoldings encourages us to apply  the local unfolding \cite{gmrr}: we first determine the smoothed level density $\rho_{sm}$ for each spectra by a histogram technique, and then integrate it numerically to obtain the unfolded eigenvalues $r_n=\int_{-\infty}^{e_N} \rho_{sm} \; {\rm d}e$. (Note, this is similar to the Gaussian unfolding except now the smoothing of $\delta$-functions is done ove the bins of fixed size).  Testing of this unfolding procedure for two fluctuation measure of a GOE, namely, the nearest neighbor spacing distribution and the number variance gives results consistent with the GOE-theory, for both bulk as well as in the edge regime.  Note, due to spurious fluctuations introduced by finite size-effects, all the unfoldings mentioned above are unreliable for large spectral length-scales. More clearly, the numerical calculation of the short-range fluctuation measure is robust across different unfolding procedures but long-range measure are sensitive  to the type of unfolding used.  Only way to distinguish the "true" fluctuations (those present in $N\rightarrow \infty$ limit) from the spurious ones is to analyze the statistics for many system-sizes and study the emergent behavior in large size limit \cite{bg}.

The fluctuations  being  in general sensitive to the spectral-range (due to non-stationarity of the spectrum), it is  necessary to choose an optimized range  $\Delta E$ which is sufficiently large for good statistics but keeps  mixing of different statistics at minimum.  We analyze  2-5${\% }$ of 
the total eigenvalues taken from a range $\Delta E$, centered at the energy-scale of interest (edge or bulk). This gives 
approximately $10^5$ eigenvalues and their eigenfunctions for each ensemble. (Note due to almost constant level density in the bulk, the statistics is locally stationary and one can take levels within larger energy ranges without mixing the statistics. A rapid variation of $\rho_{sm}$ in the edge however allows one to consider levels only within very small spectral ranges.  For edge-bulk comparisons, it is preferable to choose same number of levels for both spectral regimes). 

The standard tools for the fluctuations analysis, both experimentally as well as numerically, are  the  nearest-neighbor spacing distribution $P(s)$ and the number-variance $\Sigma^2(r)$, the measures for the short and long-range spectral correlations, respectively \cite{haak, brody}. These can be defined as follows: 
{\it {$P(s)$} (same as $P(1;s,e)$ defined in section II.A) is the probability of two nearest neighbor eigenvalues to occur at a distance $s$, measured 
in the units of local mean level spacing $D$}, and,
{\it $\Sigma^2(r)$ is the variance in the number of levels in an interval of length $r D$}. 
Figure 7 compares  $P(s)$ for Gaussian and bimodal disorders for various dimensions for fixed size $N=10^3$ and for two 
energy-ranges i.e the edge and bulk of the spectra; it clearly shows the sensitivity of $P(s)$ to both  dimensionality as well as the energy-range but almost no effect of the nature of disorder i.e Gaussian or bimodal. (note, as figures 7(a,d) indicate, the nature of disorder seems to have some influence only for $d=1$). For lower dimension or in the edge region, the statistics is shifted more towards Poisson, an indicator of the increasing eigenfunction localization.  For $d=1$, the $P(s)$ behavior seemingly approaches singularity at $s=0$, implying a high degree of quasidegeneracy among eigenvalues. As, for $d=1$, the system is nearly integrable, the singular behavior is in qualitative accordance with Shnirelman theorem which states that for a classically near integrable system at least each second level spacing in the corresponding system become exponentially small \cite{chir}. (More formally, the theorem states that the spectrum is asympotically multiple i.e for each $m >0$ there exists $C_m >0$ such that min($e_k-e_{k-1}, e_{k+1}-e_k) < C_m e_k^{-m}$ \cite{chir}).

To analyze size-dependence of the short range correlations, we 
consider $P(s)$ behavior for four different sizes of $d=\infty$  
(figure 8) and $d=3$ cases (figure 9). Although  sensitive for small values, the distribution shows a tendency to become size-independent  for large $N$-values. As the figures indicates, the limiting distribution  for both dimensions is independent 
of the nature as well as strength of disorder; this is in accord with our theoretically claimed Gaussian-bimodal analogy discussed in section 4 of \cite{ss} (between cases II and IV for $d=\infty$ and cases III, V for finite $d$).

As mentioned above, the long-range fluctuations around the spectral density can be analyzed by $\Sigma^2(r)$.  Figures 10, 11 show 
$\Sigma^2(r)$ behavior of the CCE spectra for both bulk and edge regions,  of $d=\infty$ as well as $d=3$ cases, respectively; the figures 
reconfirm the behavior indicated by $P(s)$-numerics discussed above i.e independence from the nature or strength of disorder 
but sensitivity to basis-connectivity and/ or energy-range.  
The $\Sigma^2(r)$ numerics also indicates large $N$-approach to size-invariant behavior for both $d=\infty, 3$ cases but the convergence is slower in the edge region and for large $r$-values. (Note, 
the curves for $N=512$ in figure 10 and figure 11 seem to display a non-increasing behavior but as mentioned in  section II.C, the long range measures contain spurious fluctuations for small system sizes and true fluctuations emerge only in infinite size limit \cite{bg}.  The numerics for long range measures therefore can not be considered reliable by itself however it can be used as a support for the theoretical results or for other measures (e.g. $P(s)$). As clear from figures 10,11, the curves for large N not only show smaller deviation from each other (e.g. $N=2197$ and $N=4913$) (indicating approach to size-independence as predicted theoretically by us) but their behavior is also consistent with corresponding $P(s)$ behavior.  Also note, due to finite size effects,  the statistical error in $\Sigma^2(r)$-analysis increases with large $r$).

\section{CCE-BE analogy}

The theoretical analysis discussed in section IV of \cite{ss} indicates that 
 a $N \times N$ CCE of real-symmetric matrices $H$ is analogous to a section of the BE i.e the sub-ensemble with  uniform eigenvector and zero eigenvalue.  As eq.(72) of \cite{ss} indicates, the local spectral statistics of a $N \times N$ CCE  approaches, in large $N$ limit,  the BE-statistics. 
 The ensemble density for the BE analog  can be given as 
\begin{eqnarray}
\rho(H)  \propto 
{\rm exp}{\left[-\frac{\eta}{2} \; \sum_{i=1}^{N} H_{ii}^2 -  
\eta (1+c N) \sum_{i,j=1; i < j}^{N} H_{ij}^2 \right]} 
\label{be1}
\end{eqnarray} 
with $c=1$ for the analog of a CCE with $d=\infty$; (note the theoretical analysis in \cite{ss} could not exactly predict $c$-value for the BE analog of CCE, finite $d$ case). Eq.(\ref{be1}) is also known as Rosenzweig-Porter (RP) (or Porter-Rosezweig) ensemble \cite{rp}. Referring it as a BE, however, has some advantage: it gives an incentive to seek  connections among BEs and CCEs under global symmetries other than time-reversal.

As expected due to non-stationarity, the local fluctuations  of a BE analog vary along the spectrum axis. Due to relatively smaller level-spacing, the statistics in the bulk of the  BE spectrum is very close to a GOE but the edge-statistics deviates significantly from the latter \cite{pich}.  This indicates that the GOE-CCE analogy is expected to be valid only for a very small spectral range $\Delta E$ around $e=0$ \cite{ss} which becomes narrower  as the size $N$ increases. The theoretical study in \cite{yvf} however claims  that the 2-point level-density correlation $R_2(r_1,r_2)$ ({\it the probability of finding two levels at a distance $|r_1-r_2|$}) for a CCE  is analogous to that of a GOE for all energy ranges. 
To clarify this discrepancy and find the correct CCE analog,  we numerically compare the spectral CCE statistics with that of a BE and a GOE, both in the bulk as well as edge.       
Figure 12 displays the comparative behavior for four spectral measures, namely, $R_1(r)$,  $R_2(r_1,r_2)$ as well as $P(s)$ and $\Sigma^2(r)$ for CCGE, $d=\infty$ and BE, each consisting of 5000 matrices of size $N=10^3$.
As can be seen from figure 12(a), $R_1(e)$ for the $d=\infty$ CCGE case  deviates a little from its  BE  analog but their local fluctuations measured in terms of $P(s)$, $\Sigma^2(r)$ and $R_2(r)$ are analogous.  (Although $R_2(r)$ provides the same information as given by $P(s)$ and $\Sigma^2(r)$),  it is considered here to substantiate our analytical claims against the one in \cite{yvf}).  For the clarity of presentation,  the $d=\infty$ bimodal case is not shown in figure 12 but its analogy with  corresponding CCGE case (figures 7-9) implies the same with BE.  The above comparison is also repeated for CCGE, $d=3$ case; as confirmed by figure 13, its 
BE analog, obtained by a numerical search, corresponds to $\mu \approx 200 N$.

In general, the BEs represent the non-equilibrium states of transition, driven by a single parameter, between two universality classes of random matrix ensembles. But the BE analog of a CCE with infinite connectivity is independent of any parameters except the system-size $N$. In asymptotic limit $N \rightarrow \infty$, therefore, the statistical fluctuations of  a
CCE with $d=\infty$ are free of all parameters and  represent  a new universality class (different from ten standard universality classes).  (As seen in figures 8 and 10,   both $P(s)$ and $\Sigma^2(r)$ for CCE, $d=\infty$ are size-independent; note, in figure 10, an apparent size-dependence of  $\Sigma^2(r)$ for small $N$ is due to spurious fluctuations introduced by finite size effects).

\section{Conclusion}

In this paper, we have presented a  detailed numerical analysis of the spectral statistics of the column constrained ensembles (real-symmetric case with 
independent off-diagonals and  column constant $\alpha=0$ for all columns) under various system conditions. Our study clearly indicates a non-ergodic as well as non-stationary behavior of the CCE-spectrum even for infinite connectivity;  this is in contrast to the ergodic, stationary nature of GOE (which is similar to the CCGE in terms of matrix structure except for column constraints).  This implies an energy dependence of the spectral fluctuations  which is confirmed by our numerics.  
 We also find that the strength or nature of disorder has no effect on the local spectral-fluctuations.  Although the average behavior of both, level density as well as inverse participation ratio, is sensitive to the type of disorder (for finite basis-connectivity), its dependence on the disorder-strength  can be scaled out. (Note here only those types of disorder are implied which do not change the global constraint class of the ensemble). Further both the average  behavior as well as the fluctuations  are  sensitive to  other system conditions e.g. dimensionality/ basis-connectivity and energy-range. As connectivity in the basis increases, the  average 
behavior also become insensitive to nature of disorder.

In large-size limit, the spectral fluctuations in CCEs become size-independent too. As predicted in \cite{ss} for CCE case with infinite connectivity, our numerics confirms the analogy  of its fluctuations to those of a special type of critical BE, intermediate between Poisson and GOE and free of all parameters. But, for a CCE with finite connectivity, the BE analog  depends on a single parameter. This indicates a cross-over of CCE-statistics from Poisson to this special critical BE with basis-connectivity  as the transition parameter. 

The present study still leaves many questions unanswered. For example,  the real-symmetric ensembles considered here are applicable to time-reversal systems with column constraints. It would be interesting to numerically analyze the complex Hermitian ensembles with column/row constraints; these ensembles are appropriate models, for example, for Goldstone systems without time-reversal symmetry).  The presence of correlations among off-diagonals (applicable to interacting systems), varying  column constants for each column or an absence of Hermiticity (relevant for systems e.g Google matrix) are some other important effects which are yet to be probed. Another important open question is the effect of column constraints on the  eignvector fluctuations. It would also be interesting to explore the exact dependence of the CCE-statistics on the relative values of the ensemble parameters and the column constraints e.g ratio $\alpha/ \sqrt{\gamma}$ in case III of \cite{ss}. We expect to answer some of these questions in near future.

\oddsidemargin=-30pt

\begin{figure}
\centering
\includegraphics[width=1.2\textwidth,height=1.0\textwidth]{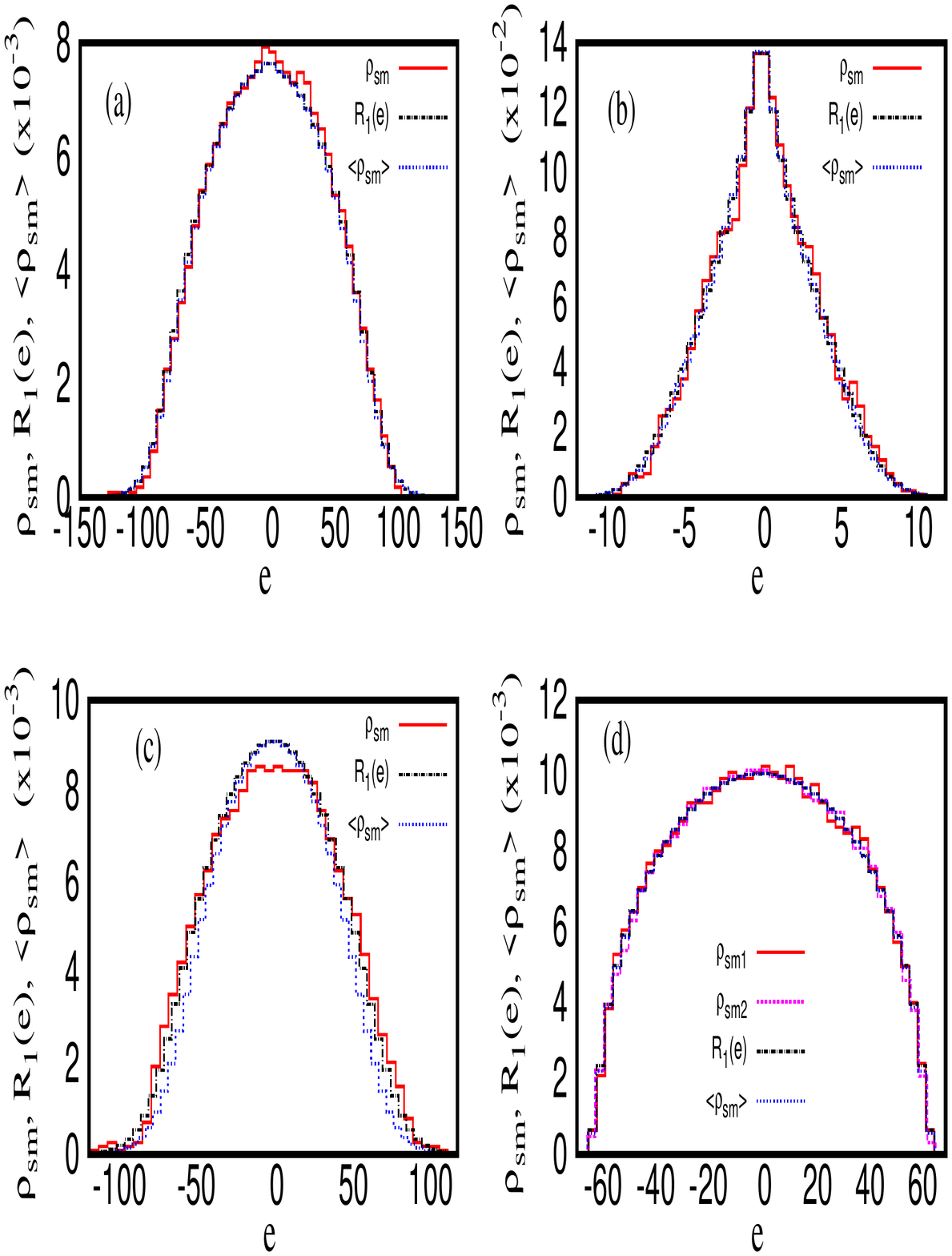} 
\caption{ 
{\bf Ergodic behavior of level density $\rho_{sm}(e)$}:  The figure compares the spectral averaged level-density $\rho_{sm}(e)$ for a single  matrix  to an ensemble average $R_1(e)$ as well as $\langle \rho_{sm}(e) \rangle$ for four ensembles: 
(a) CCGE (eq.(\ref{rho2})), (b) CCGE (eq.(\ref{rho3}), (c) BE (eq.(\ref{be1})), (d) GOE. In each case, we consider 
an ensemble of 5000 matrices of size 
$N=1000$. The numerics shows that  $\rho_{sm}$ for CCGE and BE fluctuates from one matrix to the other which manifests in 
a clear, although small,  deviation of $\langle \rho_{sm} \rangle$ from $R_1$; the deviation however seems to reduce with increasing $N$. As shown in the part (d), the 
ergodic nature of GOE is clearly visible from an exact analogy of $R_1(e)$ with $\langle\rho_{sm} \rangle$ and also with 
$\rho_{sm}$ for a single matrix (to emphasize the latter, two matrix cases are shown in figure 1(d)).    
  }
\label{fig1}
\end{figure}

\oddsidemargin=-30pt

\begin{figure}
\centering
\includegraphics[width=1.2\textwidth,height=1.0\textwidth]{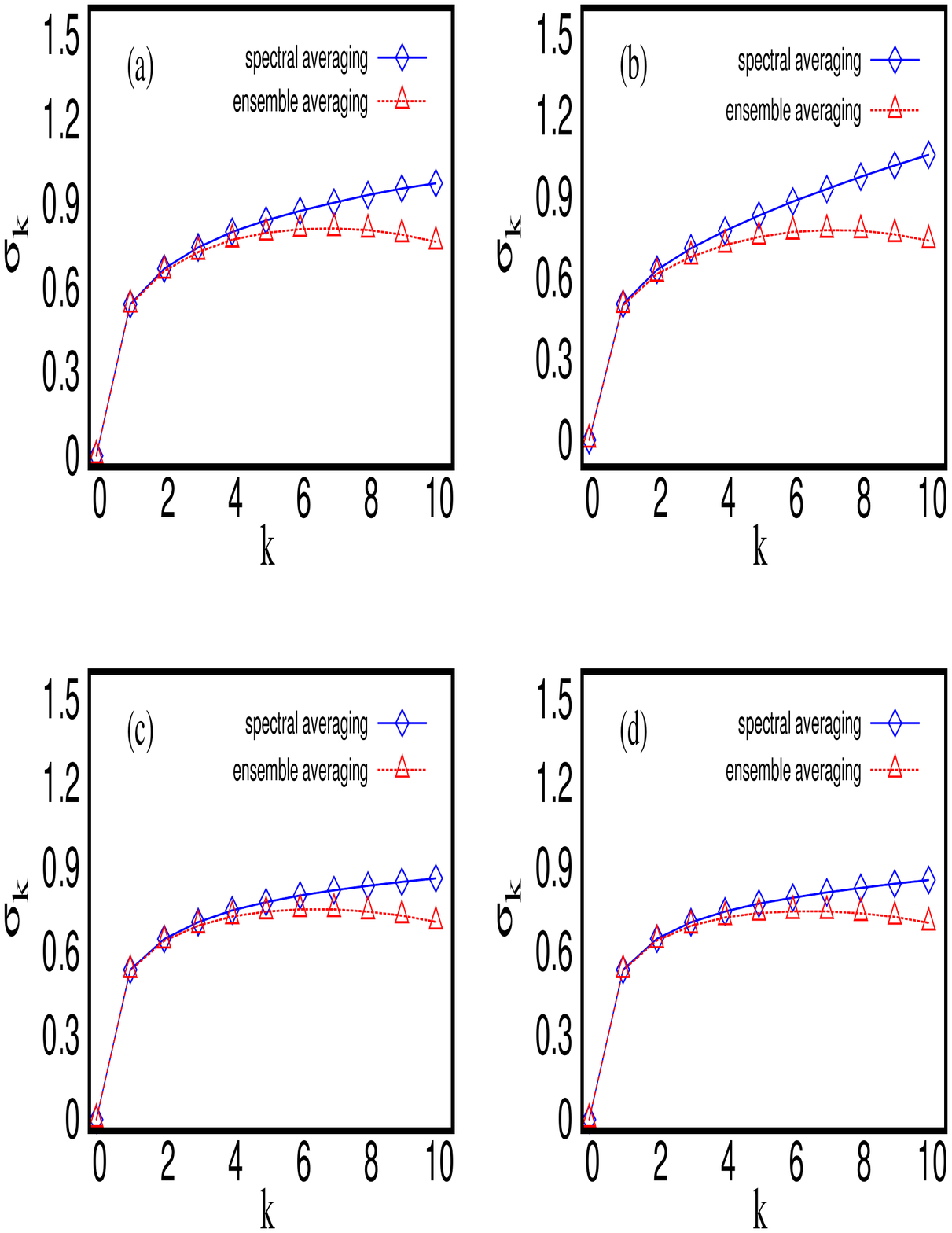} 
\caption{ 
{\bf Non-ergodicity of local fluctuations:}  The figure compares the spectral averaged  behavior of  $\sigma_k$ with ensemble average for many $k$ values and for spacings taken from bulk near $e=0$; here $\sigma_k$ measures the width  of the eigenvalue spacing distribution $P(k,x,e)$, to find $k^{th}$ nearest neighbor levels at a distance $x$ in a spectral range around $e$.  The spectral and ensemble averages are obtained by using eq.(36) and eq.(39) of \cite{bg} respectively. 
The results are shown for four different ensembles:
(a) CCGE, $d=\infty$, eq.(\ref{rho2}), (b) CCGE, $d=3$, eq.(\ref{rho3}), (c)  BE, eq.(\ref{be1}), (d) GOE. The deviation of ensemble average from the spectral one in first three case is an indicator of non-ergodicity of local fluctuations which however seems to be quite weak. The non-ergodicity is confirmed by the lack of stationarity of $\sigma_k$ shown in figure 3. Although a small deviation is seen in the case of a GOE too but it is due to spurious fluctuations and is a finite size effect; (as indicated by the stationarity displayed in figure 3(d)).  }
\label{fig2}
\end{figure}

\oddsidemargin=-30pt

\begin{figure}
\centering
\includegraphics[width=1.2\textwidth,height=1.0\textwidth]{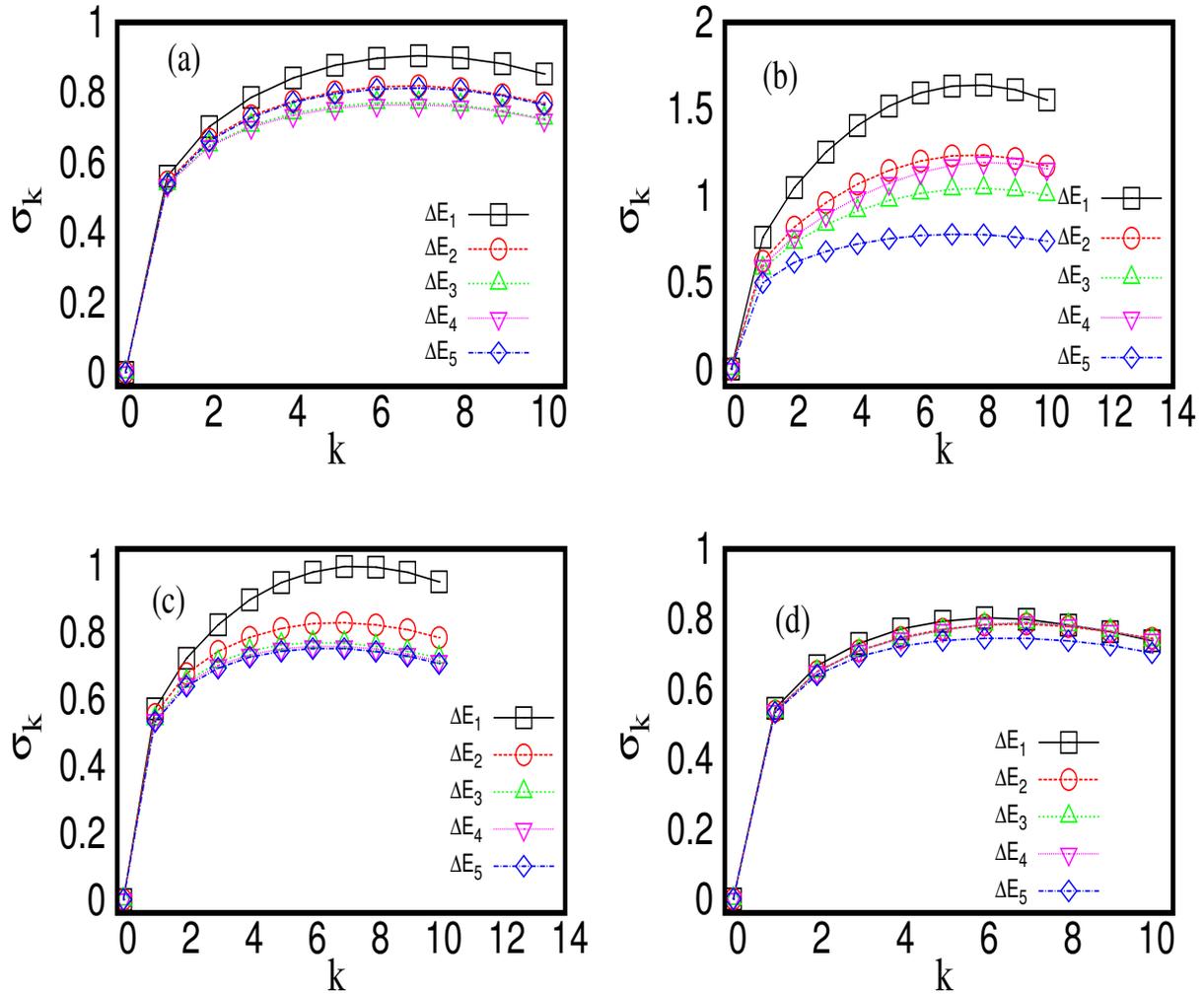} 
\caption{ 
{\bf Stationarity of local fluctuations:}  The figure displays the behavior of  ensemble-averaged $\sigma_k$, the width of the eigenvalue spacing distribution $p(k,s,e)$ for many $k$ values and for 20 level-spacings taken from the levels of five different spectral-regimes: 
(i) $\Delta E_1=98-118$, (ii) $\Delta E_2=196-216$, (iii) $\Delta E_3=294-314$, (iv) $\Delta E_4=392-412$, (v) $\Delta E_5=490-510$, 
Again four ensembles are considered:
(a) CCGE, $d=\infty$, eq.(\ref{rho2}), (b) CCGE, $d=3$, eq.(\ref{rho3}), (c)  BE, eq.(\ref{be1}), (d) GOE. Note the deviation of $\sigma_k$ for a fixed $k$ and for different spectral-ranges is significant in parts(a,b,c) while it is almost negligible in part(d); this is an indicator of the non-stationarity for the first three cases.}
\label{fig3}
\end{figure}

\oddsidemargin=-30pt

\begin{figure}
\centering
\includegraphics[width=1.2\textwidth,height=1.0\textwidth]{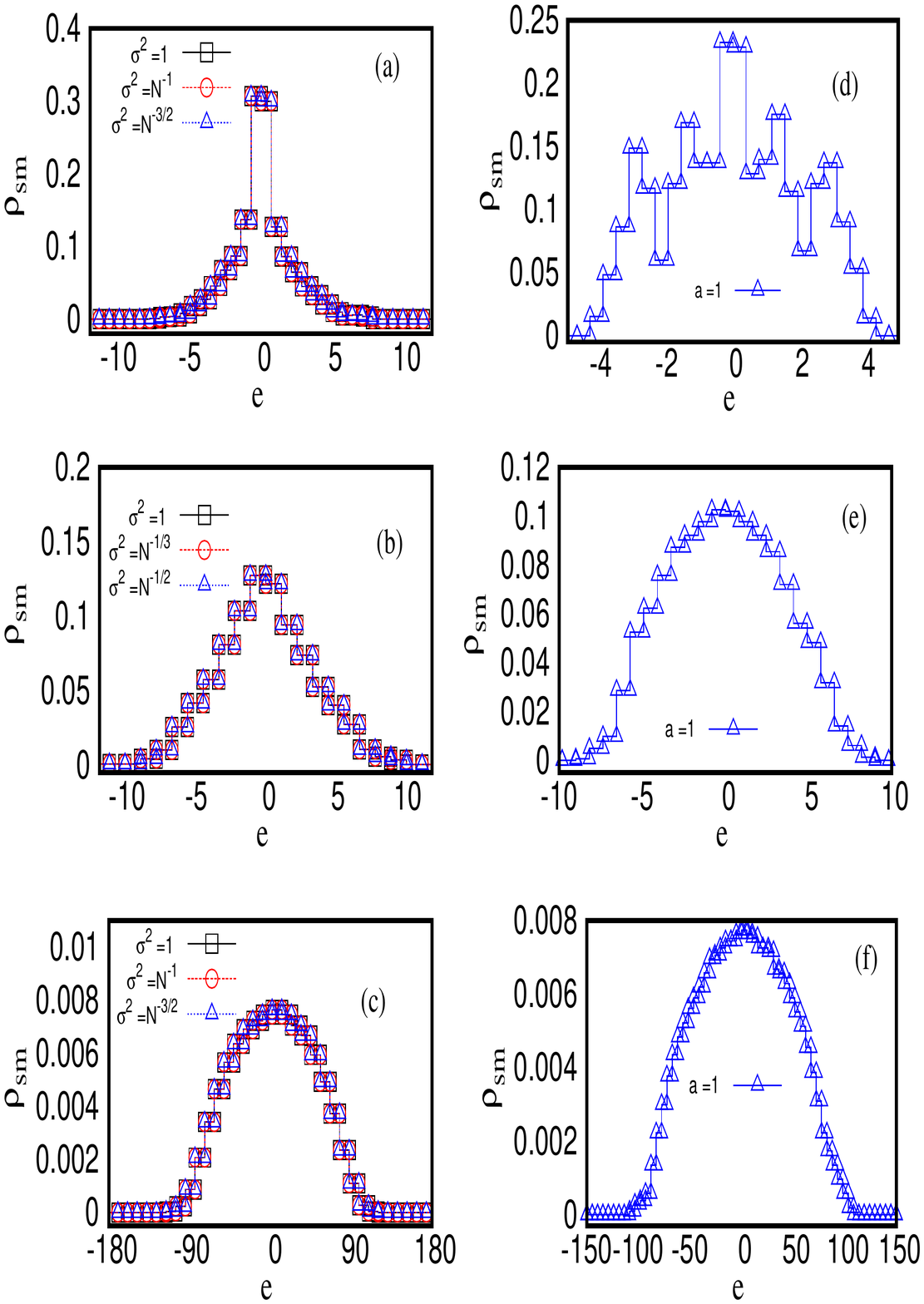} 
\caption{ 
{\bf Disorder-dependence of level density $\rho_{sm}(e)$}:  The figure illustrates the behavior of $\rho_{sm}(e)$ of a 
column-constrained ensemble  for a fixed size $N=1000$ but  for different disorder-strengths  $\gamma=\frac{1}{2\sigma^2}$ for Gaussian disorder and different basis-connectivity (to consider dimensionality effects): (a) Gaussian case $d=1$, eq.(\ref{rho3}) (b) Gaussian case $d=3$, eq.(\ref{rho3}) (c) Gaussian case, $d=\infty$, eq.(\ref{rho2}).
Note, in parts (a,b,c), $\rho_{sm}$ is scaled by $\frac{1}{N \sigma} \equiv \frac{\sqrt{2\gamma}}{N}$ which results in an analogous form for different $\sigma$;  this scaling is same as used in 
eqs.(34,54) of \cite{ss} to remove $\gamma$-dependence. To understand the dependence on the  disorder-type, we also consider bimodal case for a fixed disorder strength and three types of basis-connectivity (for $N=1000$):
 (d) bimodal case, $d=1$, eq.(\ref{birho2}) (e) bimodal case, $d=3$, eq.(\ref{birho2}) (f) bimodal case IV, $d=\infty$, eq.(\ref{birho1}).  }
\label{fig4}
\end{figure}

\oddsidemargin=-30pt
\begin{figure}
\centering
\includegraphics[width=1.2\textwidth, height=\textwidth]{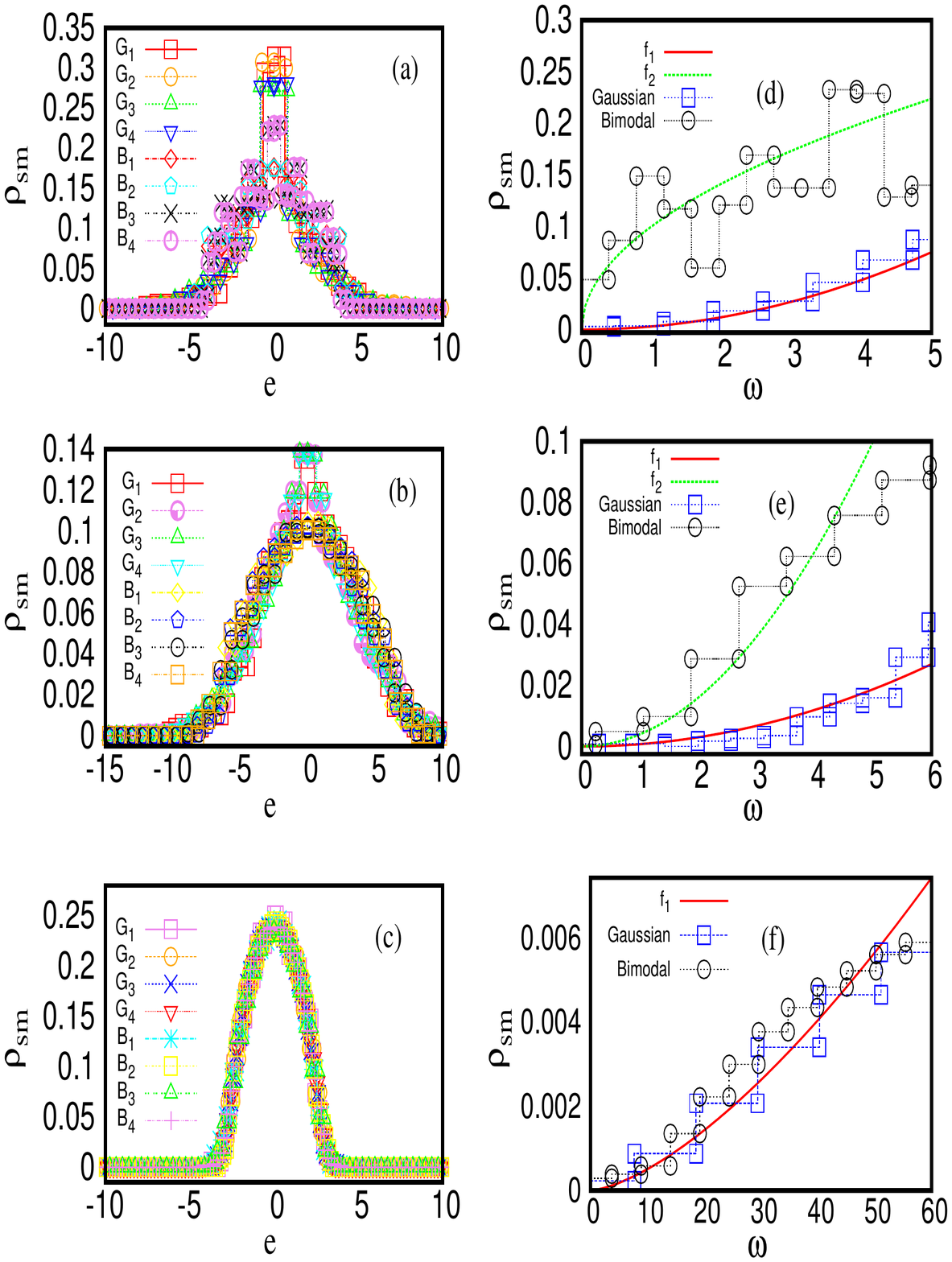} 
\vspace*{-30 mm}
\caption{
{\bf Size-dependence of level density $\rho_{sm}(e)$}:  figures (a)-(c) describes the level-density behavior of a column-constrained ensemble for a fixed disorder-strength $\sigma^2=1$ ( $\gamma=(2 \sigma^2)^{-1})$) and for different sizes $N$: (a) $d=1$, Gaussian case eq.(\ref{rho3}), bimodal case eq.(\ref{birho2}), (b) $d=3$, Gaussian case eq.(\ref{rho3}) and bimodal case eq.(\ref{birho2}), (c) $d=\infty$,  Gaussian case eq.(\ref{rho2}) and bimodal case eq.(\ref{birho1}). To avoid cluttering, the notations in (a)-(c) are changed; the symbols $G_1, G_2, G_3, G_4$  refer to  Gaussian cases with $N=512, 1000, 2197, 4913$ respectively and symbols $B_1, B_2, B_3, B_4$ refer to Bimodal cases with $N=512, 1000, 2197, 4913$ respectively. Note in parts(a,b) $\rho_{sm}$ is rescaled: $\rho_{sm} \rightarrow \rho_{sm}/N$ but in part (c), the analogy of different $N$-cases results only after following rescaling: $ e \rightarrow e/\sqrt{N},  \rho_{sm} \rightarrow \rho_{sm}/ \sqrt{N}$.  As shown later in figure 12(a), $\rho_{sm}$ for $d=\infty$ case behaves as a semi-circle in the bulk: $\rho_{sm}(e)=\frac{1}{2.3 \pi}\sqrt{4 N -e^2}$. 
The parts (d)-(f) depict the density  of the low-energy excitations for
different dimensions and two disorder-types, along with fits $f_1$ for Gaussian and $f_2$ for bimodal case: 
(d) $d=1$, $f_1= 0.003 \; x^2$, $f_2=0.01 x^{0.5}$, (e) $d=3$, $f_1=7.5 \times 10^{-4} \; x^2$, $f_2=0.004 \; 
x^2$, (f) Infinite range case, $f_1=f_2=1.6 \times 10^{-5} x^{3/2}$. The fits in figures (e)-(f) seem to 
agree with available results for Goldstone modes in spin glasses.}
\label{fig5}
\end{figure}

\oddsidemargin=-30pt
\begin{figure}
\centering
\includegraphics[width=1.2\textwidth, height=\textwidth]{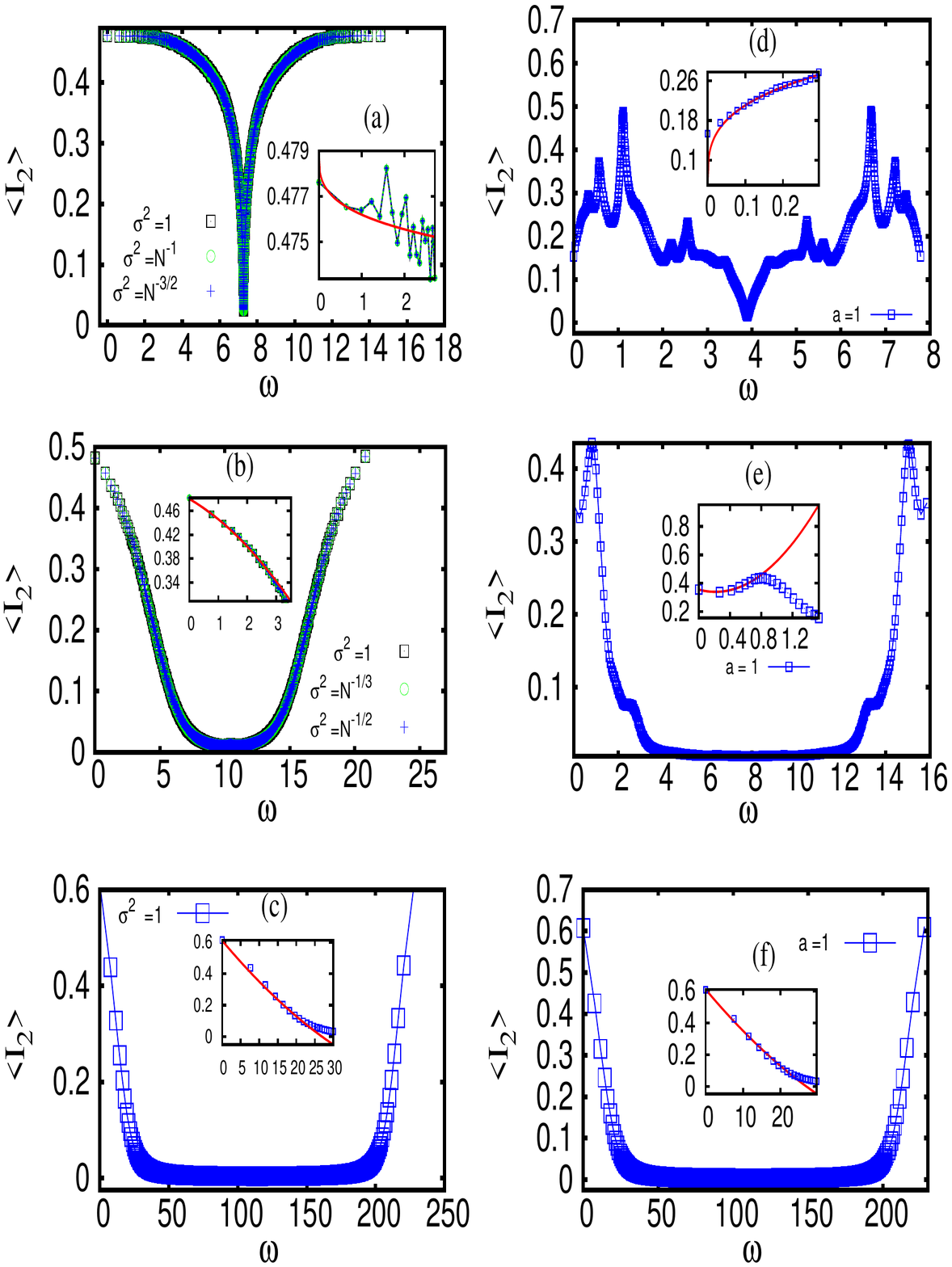} 
\vspace*{-30 mm}
\caption{
{\bf System-dependence of inverse participation ratio $I_2$}:
The figure describes the  $I_2$-behavior of a column-constrained ensemble ($N=1000$) for many disorder 
strengths (with $\gamma=(2\sigma^2)^{-1}$) (only for Gaussian disorder), basis-connectivity and for two different type of disorders. The insets in each case show the behavior for small excitation-energy $\omega$ along with the fits by solid lines:
(a) Gaussian case eq.(\ref{rho3}), $d=1$, $fit=0.479-0.0028 x^{0.3}$,
(b) Gaussian case eq.(\ref{rho3}), $d=3$, $fit= 0.485-0.028 x^{2/3}-0.01 x^2$,
(c) Gaussian case eq.(\ref{rho2}), $d=\infty$, $fit= 0.61-0.025 x + 5 \times 10^{-6} \; x^3$,
(d)  bimodal case eq.(\ref{birho2}), $d=1$, $fit=0.05+0.32 x^{0.3}$,
(e) bimodal case eq.(\ref{birho2}), $d=3$, $fit= 0.355-0.1 x^{2/3}+0.3 x^2$
(f)  bimodal case eq.(\ref{birho1}), $d=\infty$, $fit= 0.61-0.028 x+2 \times 10^{-4} \;  x^2$.
The plots in (a, b, d, e) are rescaled by the off-diagonal variance $\sigma^2$. The infinite-range case for Gaussian is considered for a single disorder-strength only; (as clear from eq.(\ref{rho2}), the off-diagonal variance for this case can  be scaled out).}
\label{fig6}
\end{figure}

\oddsidemargin=-30pt
\begin{figure}
\centering
\includegraphics[width=1.2\textwidth, height=\textwidth]{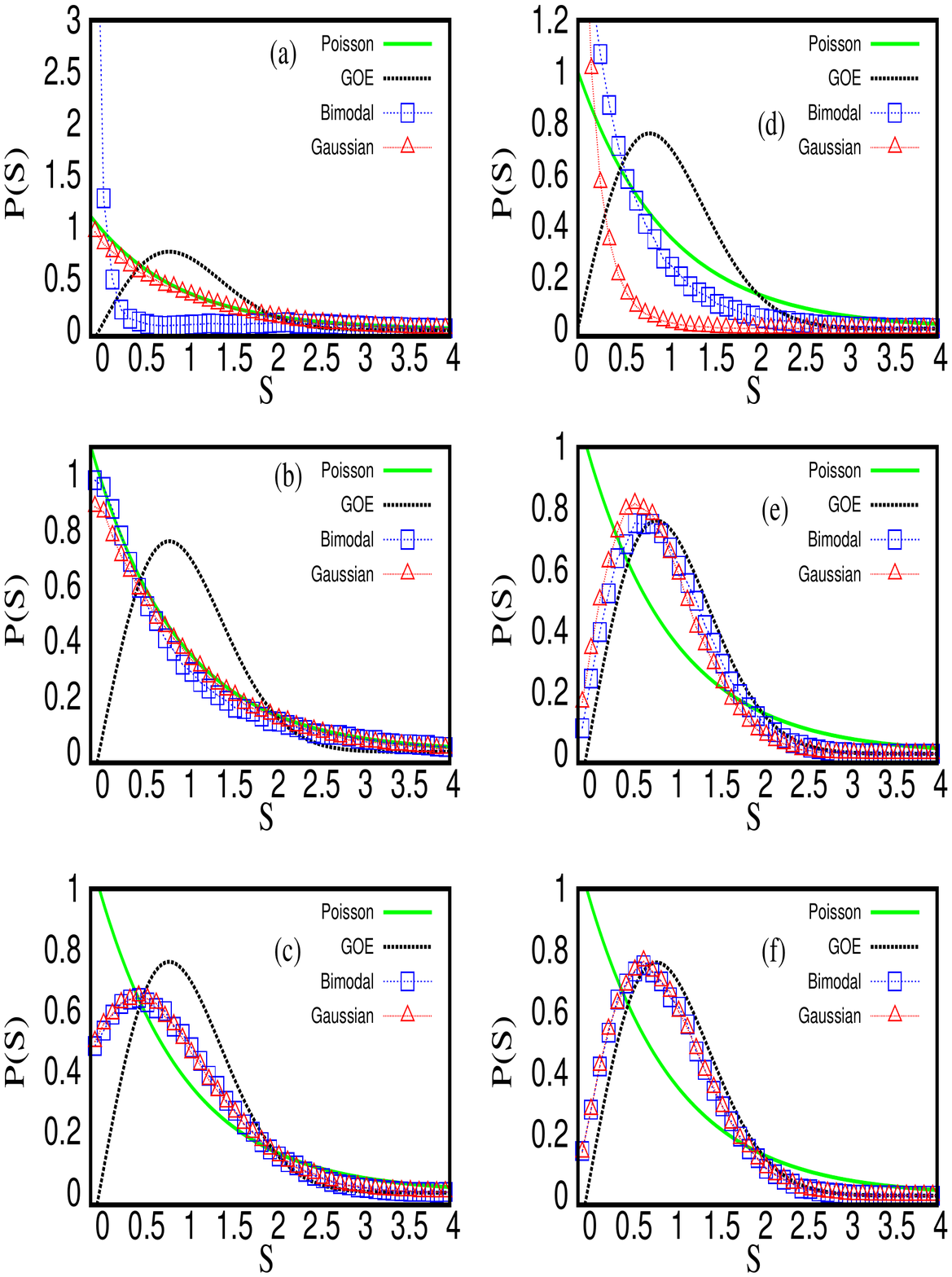} 
\vspace*{-20 mm}
\caption{
{\bf Dimensionality-dependence of nearest-neighbor spacing distribution}:
 The figure compares $P(s)$-behavior for two different types of  disorders,  and three types of basis-connectivity of the Column constrained ensemble in two different energy ranges. The size is kept fixed at $N=1000$; note, $N=L^d$ for cases with  $d=1, 3$.   
The Gaussian and bimodal cases  for $d=3$ corresponds to eq.(\ref{rho3}) and eq.(\ref{birho2}) respectively .  For $d=\infty$,  
eq.(\ref{rho2})  and eq.(\ref{birho1})  give  the Gaussian and bimodal cases respectively.  
The parts (a)-(c) describe the statistics in the edge and (d)-(f) in the bulk of the spectra:
(a) edge,  $d=1$, 
(b) edge,  $d=3$,
(c) edge, $d=\infty$, 
(d) bulk,  $d=1$, 
(e) bulk,  $d=3$
(f) bulk,  $d=\infty$. 
The figures  suggest an increase of level-repulsion with increasing basis-connectivity, with rate of change slower in the edge than bulk. The difference in the edge and bulk statistics for each dimensionality also indicates a lack of stationarity in the spectrum.}
\label{fig7}
\end{figure}

\oddsidemargin=80pt
\begin{figure}
\centering
\includegraphics[width=1.2\textwidth, height=0.8\textwidth]{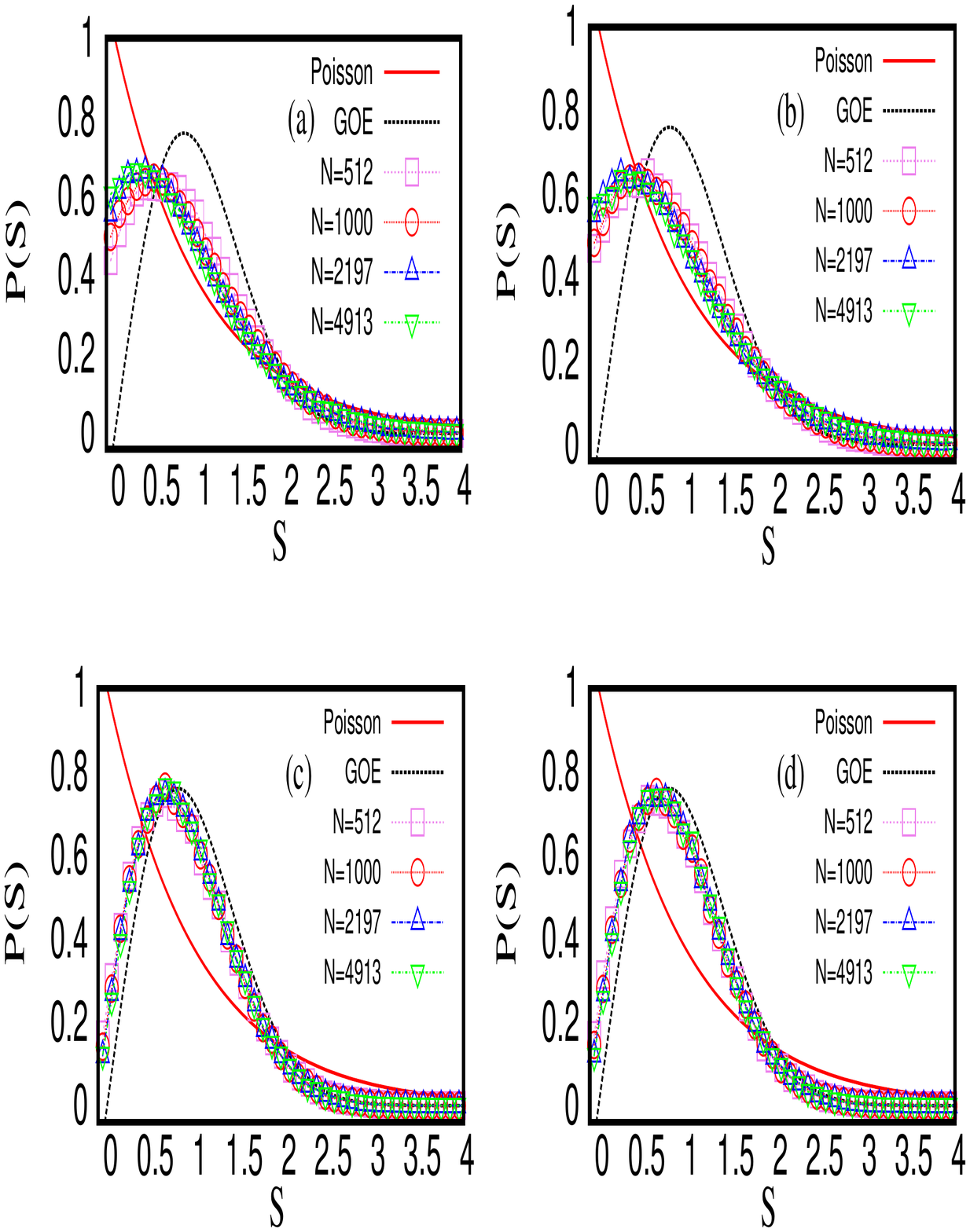} 
\caption{
{\bf Size-dependence of  $P(s)$ for column constrained ensemble with $d =\infty$}:
The figure describes the $P(s)$-behavior for the Gaussian case  (eq.(\ref{rho2})  and bimodal case (eq.(\ref{birho1})) for four system sizes $N$.  As seen in the figure, $P(S)$ approaches to an  invariant form as the system size $N$ increases. 
The limit is  sensitive to the energy-range (e.g. edge vs bulk) but almost independent of the nature of disorder (e.g. Gaussian vs bimodal) :(a) edge, Gaussian, (b) edge, bimodal, (c) bulk, Gaussian, (d) bulk, bimodal. Note  the statistics in the bulk is close to a GOE as expected for a BE with $\mu= N$ (only $2 \%$ levels chosen from the center of the bulk in each case). 
In the edge regime, intermediate behavior of $P(s)$ to  Poisson and GOE limits is again in agreement with theoretically expectation (as $\Lambda_{bulk} \sim 1 > \Lambda_{edge}$ (see section V.B of \cite{ss})).}
\label{fig8}
\end{figure}

\oddsidemargin=30pt
\begin{figure}
\centering
\includegraphics[width=1.2\textwidth, height=0.8\textwidth]{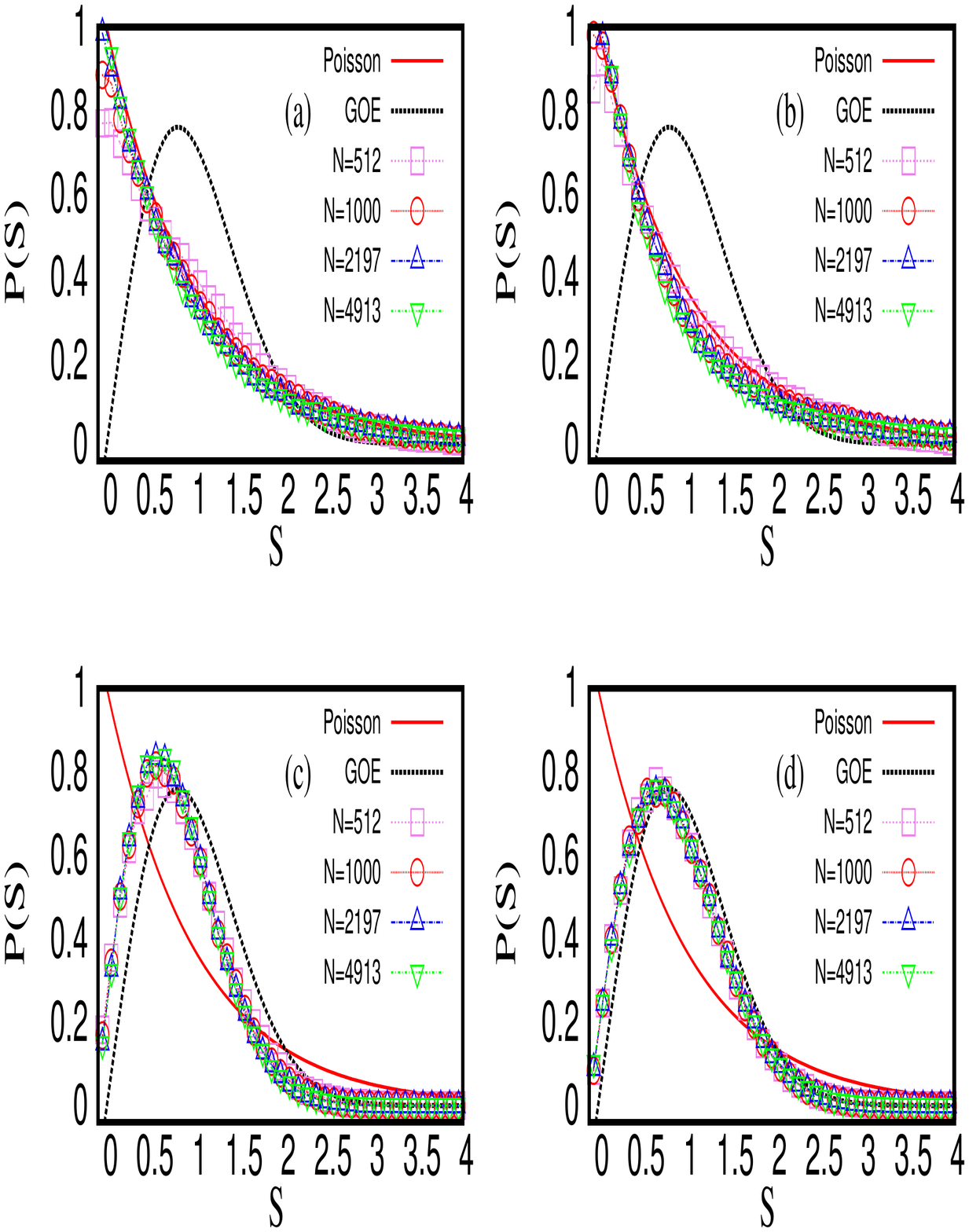} 
\caption{
{\bf Size-dependence of  $P(s)$ for $d=3$ case}:
The figure shows the $P(s)$-behavior for Gaussian case (eq.(\ref{rho3}) 
and bimodal case V  (eq.(\ref{birho2}) for four sizes $N$ for the case $d=3$ where $N=L^3$ and
periodic boundary conditions are imposed at length $L$.  Here again the approach of $P(S)$ to an invariant form with increasing $N$ is indicated.  Similar to $d=\infty$ case, here also the approach is  sensitive to the energy-range (e.g. edge vs bulk) but is almost independent of the nature of disorder (e.g. Gaussian vs bimodal) :
(a) edge, Gaussian, (b) edge, bimodal, (c) bulk, Gaussian, (d) bulk, bimodal. As compared to $d=\infty$ case (fig 8), the edge-statistics here is shifted more towards Poisson regime which is expected due to an increased localization of the eigenfunctions, originating in confined hopping. }
\label{fig9}
\end{figure}

\oddsidemargin=-10pt
\begin{figure}
\centering
\includegraphics[width=1.2\textwidth, height=0.8\textwidth]{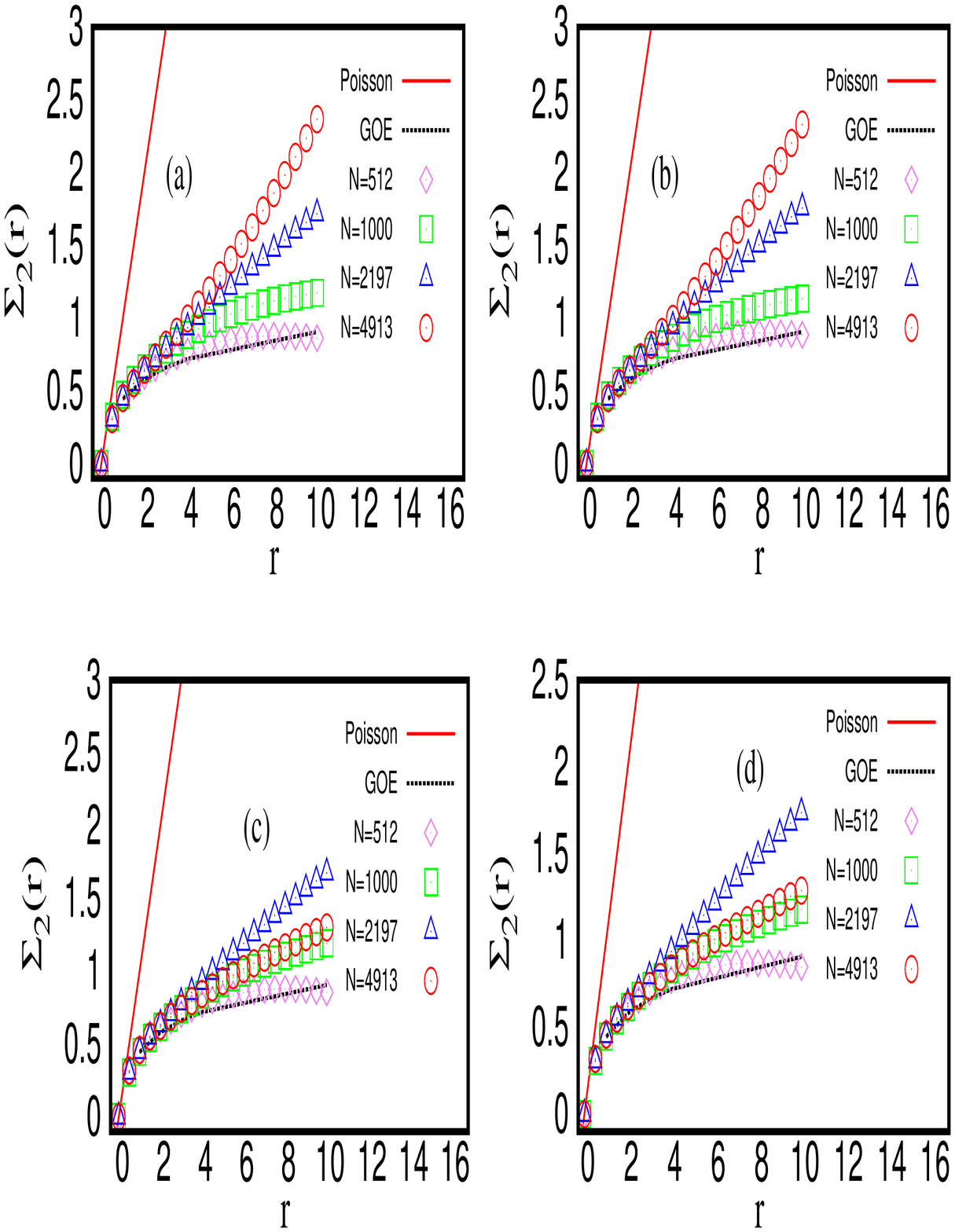} 
\caption{ {\bf  Size-dependence of the number-variance $\Sigma_2(r)$ for case $d=\infty$}: figure shows the variance of number of levels in a distance of $r$ mean level spacings for  Gaussian case (eq.(\ref{rho2})), and bimodal case (eq.(\ref{birho1})):
 (a) Gaussian, edge, (b) Bimodal, edge, (c) Gaussian, bulk, (d) Bimodal, bulk. To detect the $N$-sensitivity  in $N \rightarrow \infty$ limit, each case is considered for four system sizes. As visible in  parts 10(c,d), the statistics in the bulk is again near a GOE and reconfirms  $P(s)$ behavior shown in figure 8(c,d). Note, the approach to size-invariance for $\Sigma^2(r)$ in the edge region is slower for large $r$ but is visible from the behavior of curves for $N=2197, 4913$ in parts 10(a,b). The statistics now is intermediate between Poisson and GOE limits; this is in conformity with corresponding $P(s)$ behavior shown in figure 8(a,b).}
\label{fig10}
\end{figure}

\oddsidemargin=-10pt
\begin{figure}
\centering
\includegraphics[width=1.2\textwidth, height=0.8\textwidth]{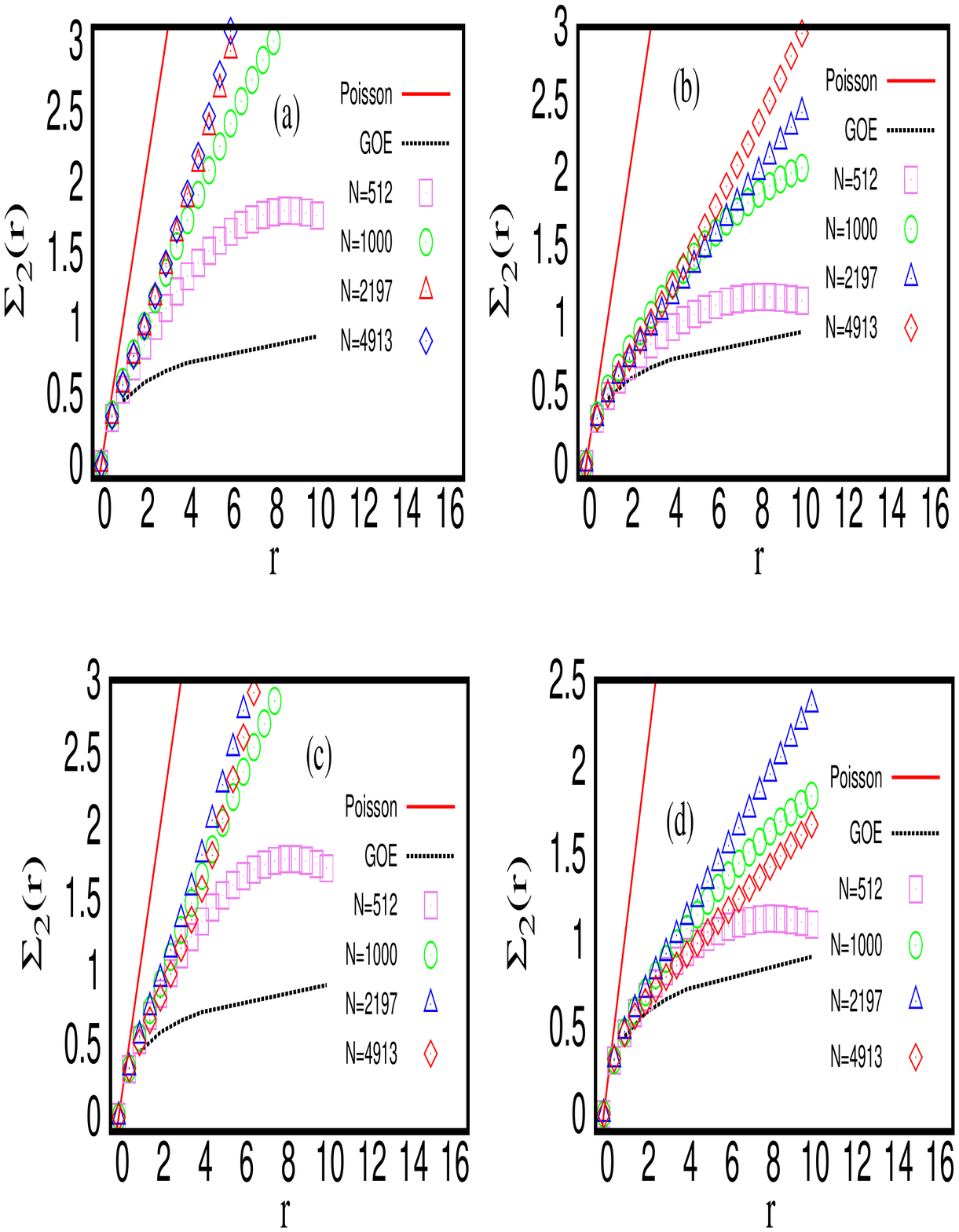} 
\caption{
{\bf Size-dependence of number-variance $\Sigma_2(r)$ for $d=3$ case}:
The Figure shows the variance of number of levels in a distance of $r$ mean level spacings for the Gaussian case (eq.(\ref{rho3})), and, the bimodal case (eq.(\ref{birho2})  for four sizes $N$: (a) Gaussian, edge, (b) Bimodal, edge, (c) Gaussian, bulk, (d) Bimodal, bulk.  Here again, the approach to size-invariance in the edge region is slower for large $r$ but is visible from the behavior of large $N$-curves; similar to $P(s)$ behavior for $d=3$ (figure 9), the statistics here is intermediate between Poisson and GOE limits.}
\label{fig11}
\end{figure}

\oddsidemargin=-10pt
\begin{figure}
\centering
\includegraphics[width=1.2\textwidth, height=1.4\textwidth]{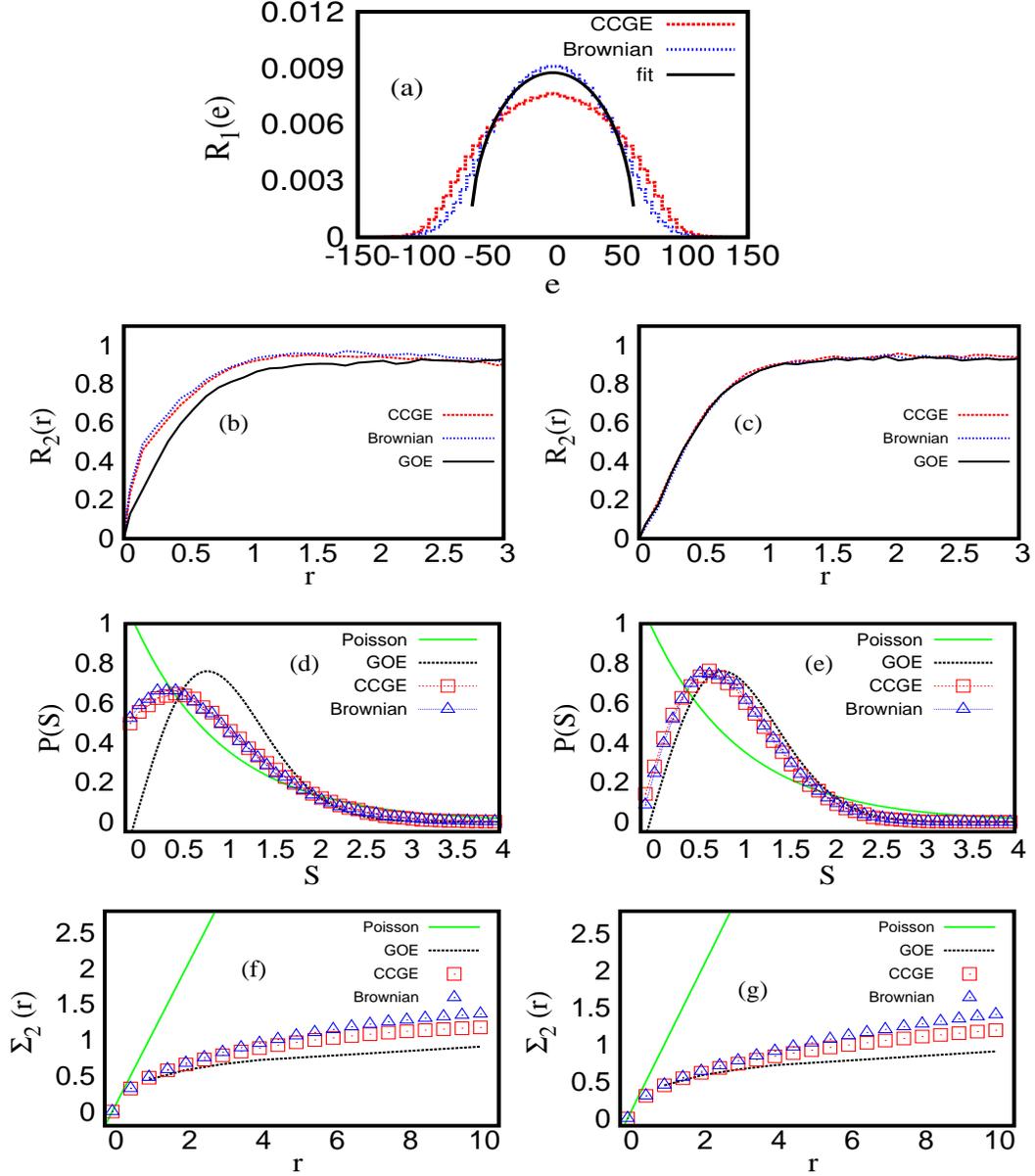} 
\vspace*{-65 mm}
\caption{ 
{\bf Comparison of column constrained ensemble, $d=\infty$ with BE}: 
Here we compare CCGE given by eq.(\ref{rho2}) along with its BE analog (given by eq.(\ref{be1}) with $\mu=N$). Each ensemble is considered for a fixed size $N=1000$ but  the results shown are independent of the system size in the study: (a) $R_1(e) ($ rescaled by $N$), (b) $R_2(e)$, edge, (c) $R_2(e)$, bulk
(d) $P(s)$, edge, (e) $P(s)$, bulk, (f) $\Sigma^2(r)$, edge, (g) $\Sigma^2(r)$, bulk.
The part (a) clearly indicates the deviation, although small, of   $R_1(e)$ for  CCGE from that of BE; as expected, the semi-circle {\rm fit}=$\frac{1}{2.3 \pi N} \sqrt{4N - e^2}$ agrees well with the bulk of BE but deviates in the tail regime.  As seen in the parts (b,c), $R_2(r)$ for CCGE deviates from that of GOE in the edge but agrees well in the bulk.  
The parts (d)-(g) confirm the CCE-BE analogy of the local fluctuations for Gaussian disorder, both in the bulk as well as in the edge-region.
The same analogy is valid for CCE case with bimodal disorder too which, although not shown here for clarity, is evident from the analogous behavior of Gaussian and Bimodal cases shown in the figures 7(c,f) for $P(s)$ and figure 10 for $\Sigma_2(r)$.
}
\label{fig12}
\end{figure}

\oddsidemargin=-10pt
\begin{figure}
\centering
\includegraphics[width=1.2\textwidth, height=1.4\textwidth]{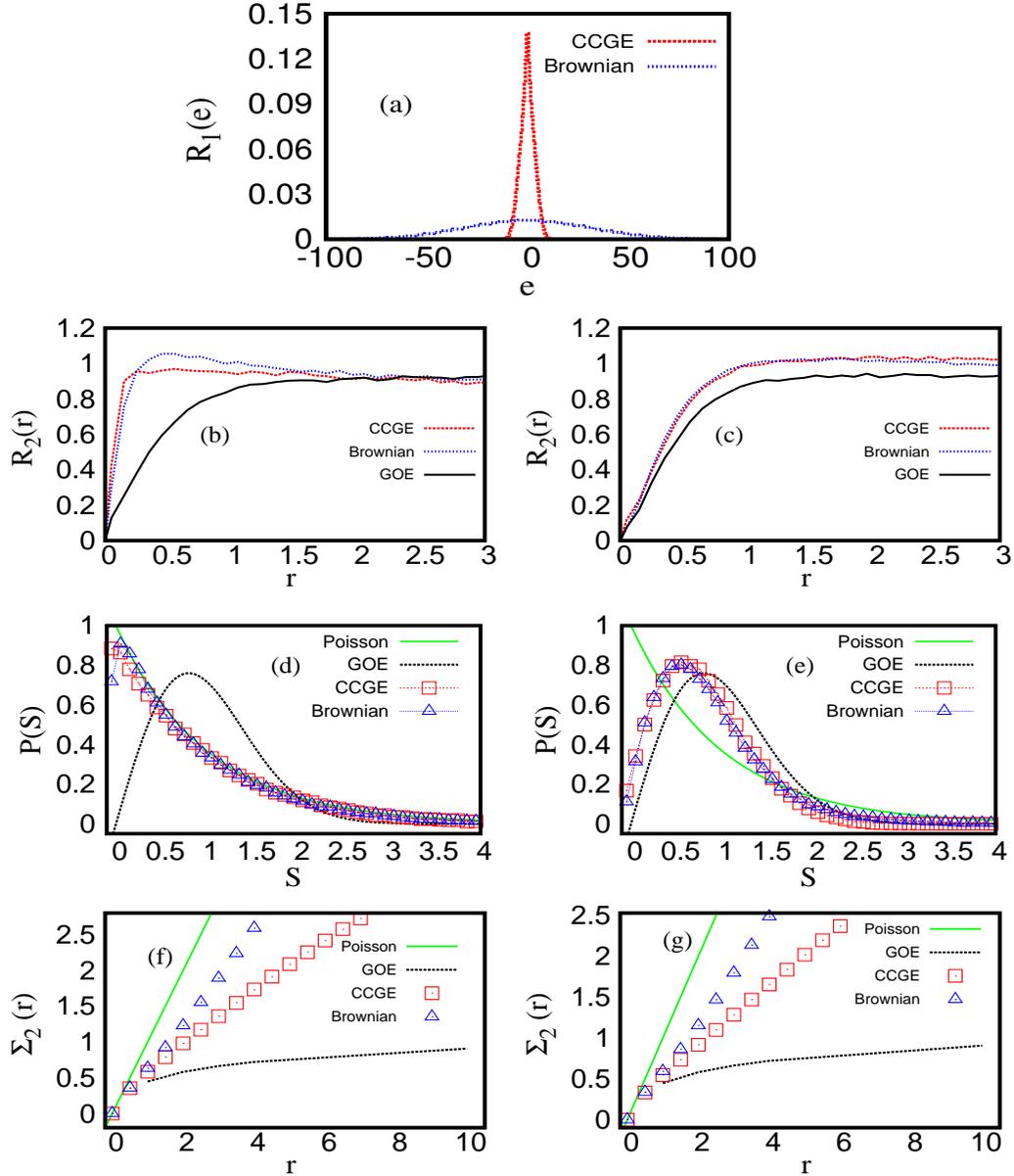} 
\vspace*{-60 mm}
\caption{ 
{\bf Comparison of column constrained ensemble, $d=3$ case with BE}: 
Here we compare CCGE given by eq.(\ref{rho3}) with the BE, given by eq.(\ref{be1}) with $c = 200$. Each ensemble is considered for a fixed size $N=1000$ and four four fluctuation measures: 
(a) $R_1(e) ($ rescaled by $N$), (b) $R_2(e)$, edge, (c) $R_2(e)$, bulk
(d) $P(s)$, edge, (e) $P(s)$, bulk, (f) $\Sigma^2(r)$, edge, (g) $\Sigma^2(r)$, bulk.
The deviation of $R_1(e)$ for the CCGE from that of BE is now clearly visible from the part(a). But, as the parts(b, c, d, e) indicate,  $R_2$ and $P(s)$ in both bulk and edge agree well for the two cases.  A small deviation seen in the parts (f,g) for $\Sigma_2(r)$ may be a finite size-effect. Another possibility is that the BE with $c=200$ is although close but is not an exact analog of the CCGE case considere here. Note our theoretical analysis given in section IV.B of \cite{ss} does not exactly predict the BE analog for this case. 
}
\label{fig13}
\end{figure}

\end{document}